\documentclass[hyper,letterpaper,12pt]{article}
\usepackage{a4wide}
\usepackage{amsmath}
\usepackage[
      colorlinks=true,
      linkcolor=blue,
      urlcolor=blue,
      filecolor=black,
      citecolor=blue,
            ]{hyperref}
\usepackage{color}
\usepackage{graphicx}
\usepackage{amsfonts}
\usepackage{amssymb}
\usepackage{epsfig}
\usepackage{subfigure}
\usepackage{amsmath}
\usepackage{latexsym}
\usepackage{mathrsfs}
\newcommand{\be}{\begin{equation}}
\newcommand{\ee}{\end{equation}}
 \newcommand{\bea}{\begin{eqnarray}}
 \newcommand{\ena}{\end{eqnarray}}

\begin{document}
\title{\bf \Large  Shadows and photon spheres with spherical accretions in the  four-dimensional  Gauss-Bonnet black hole}

\author{\large
~~Xiao-Xiong Zeng$^{1,2}$\footnote{E-mail: xxzengphysics@163.com},
Hai-Qing Zhang$^{3,4}$\footnote{E-mail: hqzhang@buaa.edu.cn},
Hongbao Zhang$^{5,6}$\footnote{E-mail: hzhang@vub.ac.be}
\date{\today}
\\
\\
\small $^1$ State Key Laboratory of Mountain Bridge and Tunnel Engineering, Chongqing\\
\small Jiaotong University, Chongqing 400074, China\\
\small $^2$ Department of Mechanics, Chongqing Jiaotong University, Chongqing ~400074, China\\
\small $^3$ Center for Gravitational Physics, Department of Space Science, Beihang University,\\
\small Beijing 100191, China\\
\small $^4$ International Research Institute for Multidisciplinary Science, Beihang University, \\
\small Beijing 100191, China\\
\small $^5$  Department of Physics, Beijing Normal University, Beijing 100875, China\\
\small $^6$  Theoretische Natuurkunde, Vrije Universiteit Brussel, and The International Solvay \\
\small Institutes, Pleinlaan 2, B-1050 Brussels, Belgium\\}

\maketitle

\begin{abstract}
\normalsize
We investigate the shadows and photon spheres of the four-dimensional Gauss-Bonnet black hole with the static and infalling spherical accretions. We show that for both cases, there always exit shadows  and  photon spheres.  The radii of the shadows  and  photon spheres are independent of the profiles of accretion for a fixed Gauss-Bonnet constant, implying that the shadow is a signature of the spacetime geometry and it is hardly
influenced by accretion. Because of the Doppler effect, the shadows of the infalling accretion are found to be darker than that of the static one.  We also investigate the effect of the Gauss-Bonnet constant on the shadow and photon spheres, and we find that the larger the Gauss-Bonnet constant is, the smaller the radii of the shadow  and  photon spheres  will be.  In particular, the observed specific intensity increases as the Gauss-Bonnet constant grows.

\end{abstract}
\newpage


\section{Introduction}\label{sec1}

The Event Horizon Telescope (EHT) collaboration has recently
  obtained
an ultra high angular resolution image of the accretion
flow around the supermassive black hole in M87* \cite{Akiyama:2019cqa, Akiyama:2019brx, Akiyama:2019sww, Akiyama:2019bqs, Akiyama:2019fyp, Akiyama:2019eap}.  The image shows that there is a  dark interior with a bright ring  surrounding it. The dark interior is called black hole shadow while the bright ring is called photon ring, respectively.
The shadow  of a black hole is caused by gravitational
light deflection \cite{Synge:1966okc, Bardeen:1972fi, Gralla:2019drh, Allahyari:2019jqz,Li:2020drn}.   Specifically, when light  emitting from the accretion passes  through the vicinity of
the black hole  toward the observer,  its trajectory will be  deflected. The intensity of the light  observed by the distant observer differs accordingly, leading to a  dark   interior and   bright ring.
So far, the shadows of various  black holes have been investigated. It is generally known that the shadows  of   spherically symmetric black holes  are round and those of rotating black holes are not precisely round but
deformed.

Since the release of the image and data by EHT, its various implications have been explored. For instance,   the  extra dimensions could be determined  from the  shadow
of M87* \cite{Banerjee:2019nnj, Vagnozzi:2019apd}, where a  rotating braneworld black hole was considered.
 The shadows of high-redshift supermassive black holes may serve as the
standard rulers \cite{Vagnozzi:2020quf}, whereby the cosmological parameters can be constrained.
The black hole companion for M87* can also be constrained through the
image released by  EHT
\cite{Safarzadeh:2019imq}. Moreover, the information given by  EHT can be used to impose constraints on particle physics
via the mechanism of superradiance
\cite{Davoudiasl:2019nlo,Roy:2019esk}. In particular, for the  vector boson, it may constrain some of the
fuzzy dark matter parameter space.  In addition, dense axion cloud can also
be induced by rapidly rotating black holes through superradiance \cite{Chen:2019fsq}.

Accretion matters are apparently important for the shadows of black holes, since many astrophysical  black holes are believed to be surrounded by accretion matters.
The first image of a black hole surrounded by an thin disk accretion was pictured out in
\cite{Luminet:1979nyg}. For a geometrically and optically thick accretion disk \cite{Cunha:2019hzj}, it was found that the mass of the disk would affect the shadow of the black hole,  and
as the mass grows the shadow becomes more prolate.  In particular, by reanalyzing the trajectory of the light ray, the shadow of a Schwarzschild black hole with both thin and thick accretion disks  have been clarified and detailed recently
\cite{Gralla:2019xty}. It was found that there existed not only the photon ring\footnote{Note that in this paper, the photon ring is   defined by the light ray that   intersects the plane of the disk
three or more times, which is  different from other references such as Ref.\cite{Narayan:2019imo}. To distinguish them, we call the photon ring in Ref.\cite{Narayan:2019imo} as the photon sphere in this paper.  } but also the lensing ring. The lensing ring makes  a significant contribution to the observed
flux while the  photon ring makes little. In addition, the observed size of the central dark area was found to be determined not only by the
gravitational redshift  but also by
the emission profile.
When the accretion matter is spherically symmetric, there is also a shadow for the black hole
\cite{Falcke:1999pj}.  The location of the shadow edge is found to be  independent of the inner radius at which the accreting gas stops
radiating \cite{Narayan:2019imo}. The size of the observed shadow can serve as a signature of the spacetime geometry, since it is hardly  influenced by the details of the accretion. This result is different from the case in which the accretion is a disk \cite{Gralla:2019xty}.

In this paper, we intend to investigate the shadow of a four-dimensional Gauss-Bonnet black hole with spherical accretions \cite{Glavan:2019inb}. The Gauss-Bonnet term in the Lagrangian is topologically invariant in
four dimensional spacetime. Thus in order to consider the dynamical effect of Gauss-Bonnet gravity, one is generically required to work in higher dimensions
\cite{Cai:2001dz, Boulware:1985wk}.  Very recently, Glavan and Lin has proposed a Gauss-Bonnet modified gravity in four dimension by simply rescaling the Gauss-Bonnet coupling constant  $\alpha\to\ \alpha/(D-4)$ and
taking the limit $D\rightarrow 4$ \cite{Glavan:2019inb}. However, as many authors have pointed out, this theory is not well-defined with the initial regularization
scheme \cite{  Hennigar:2020lsl, Tian:2020nzb, Shu:2020cjw, Mahapatra:2020rds}. Recently, the authors in \cite{Aoki:2020lig} have proposed a consistent theory of four dimensional Gauss-Bonnet gravity using ADM decomposition of the spacetime. They successfully found a four dimensional Gauss-Bonnet theory of two dynamical degrees of freedom by breaking the temporal diffeomorphism invariance.  Thus, the cosmological and black hole solutions naively given in \cite{Glavan:2019inb} can be accounted as exact solutions in the theory of \cite{Aoki:2020lig}. Our background of the black hole solution indeed also satisfies the well-defined theory of \cite{Aoki:2020lig}.  Many other characteristics of the four-dimensional Gauss-Bonnet black hole have been investigated, see for instance
\cite{Konoplya:2020bxa, Konoplya:2020cbv,  Li:2020tlo, Mishra:2020gce, Zhang:2020qam, Lu:2020iav, Zhang:2020qew,Konoplya:2020qqh,  Fernandes:2020rpa, Liu:2020evp,HosseiniMansoori:2020yfj,Roy:2020dyy, Singh:2020xju, Aragon:2020qdc,Kumar:2020owy,Zeng:2020vsj}. 

In particular,
gravitational lensing by black holes in ordinary medium  and  homogeneous plasma in  four-dimensional Gauss-Bonnet gravity have been studied in \cite{Islam:2020xmy, Jin:2020emq}. The shadows cast by  the spherically symmetric \cite{Konoplya:2020bxa, Guo:2020zmf}  and rotating \cite{Wei:2020ght} four-dimensional  Gauss-Bonnet black hole have also been studied. It will be more interesting to investigate the corresponding light intensity of the shadow, which comprises the main issue of this paper. To be more precise, in this paper, we are interested in the spherical accretions, which can be classified into the static and infalling one.
On the one hand, we want to explore how the Gauss-Bonnet constant  affects the radii of the shadow and photon sphere as well as the light intensity observed by a distant observer. On the other hand, we want to explore how the dynamics of the accretion affects the shadow of the black hole.  As a result, we find that the larger the Gauss-Bonnet constant is, the smaller the   radii of the shadow  and  photon sphere  will be, and the larger the intensity will be.  In addition, the shadow of the infalling accretion are found to be darker than that of the static case because of the Doppler effect.

The remainder of this paper is organized as follows. In Section 2, we investigate the motion of the light ray near the four-dimensional Gauss-Bonnet black hole and figure out  how it is deflected.  In Section 3, we  investigate the shadows and photon spheres with the static spherical accretion. To explore whether the dynamics of the accretion will affect the  shadow and photon sphere,  the accretion is supposed to be infalling in Section 4.  Section 5 is devoted to the conclusions and discussions. Throughout this paper, we set $G=\hbar=c
=  k_B=1$.

\section{Light deflection in the four-dimensional Gauss-Bonnet black hole }\label{sec2}
Starting from the following   Einstein-Hilbert action with an additional Gauss-Bonnet term
\be
I=\frac{1}{16\pi G}\int\sqrt{-g} d^4x\left[R+\alpha(R_{\mu\nu\lambda\delta}R^{\mu\nu\lambda
\delta}-4R_{\mu\nu}R^{\mu\nu}+R^2)\right],\label{action}
\ee
by rescaling the Gauss-Bonne coupling constant $\alpha\to\ \alpha/(D-4)$ and taking the limit $D\to4$, one can obtain the four-dimensional spherically symmetric Gauss-Bonnet black hole  as
\be\label{metric}
ds^2=-F(r)dt^2+\frac{dr^2}{F(r)}+r^2(d\theta^2+\sin^2\theta d\phi^2),
\ee
with
\be
F(r)=1+\frac{r^2}{2\alpha}\left(1-\sqrt{1+\frac{8\alpha M}{r^3}}\right),\label{fr}
\ee
where $M$ is the mass of the black hole. Note that  the same solution was already found previously in
\cite{Cai:2009ua}  by considering  the Einstein gravity with Weyl anomaly.  Solving the equation $F(r)=0$, one can
obtain two solutions,
\be
r_\pm=M\pm\sqrt{M^2-\alpha},\label{horizon}
\ee
in which $r_+$ and  $r_-$ correspond to the outer horizon (event horizon) and inner horizon, respectively. In order  to assure the existence of a horizon, the Gauss-Bonnet coupling constant should be restricted in  the range $-8\leq\alpha/M^2\leq1$. For the case $\alpha>0$ there are two horizons, while for the case $\alpha<0$ there is only one single horizon.

In order to investigate the light deflection  caused by  the four-dimensional Gauss-Bonnet black hole, we need to find how the light ray moves around the black hole. As we know, the light ray satisfies the geodesic equation, which can be encapsulated in the following Euler-Lagrange equation
\begin{equation}
\frac{d}{d\lambda}\left(\frac{\partial \mathcal{L}}{\partial \dot{x}^{\mu}}\right)=\frac{\partial \mathcal{L}}{\partial x^{\mu}},
\label{me}
\end{equation}
with $\lambda$ the affine parameter, $\dot{x}^{\mu}$ the four-velocity of the light ray and $\mathcal{L}$ the Lagrangian, taking the form as
\begin{equation}
 \mathcal{L}=\frac{1}{2}g_{\mu\nu}\dot{x}^{\mu}\dot{x}^{\nu}=\frac{1}{2}\left( - F(r)\dot{t}^2+\frac{\dot{r}^2}{F(r)}+r^2\left(\dot{\theta}^2+\sin^2{\theta}~ \dot{\phi}^2 \right)\right).
\label{la}
 \end{equation}
 As in \cite{Synge:1966okc, Bardeen:1972fi, Gralla:2019drh}, we focus on the light ray that moves on the equatorial plane, i.e., $\theta=\frac{\pi}{2}$ and
$\dot{\theta}=0$.  In addition, since none of the metric coefficients depends explicitly on time $t$ and azimuthal angle $\phi$, there are two corresponding conserved quantities, $E$ and $L$.
  Combining Eqs.(\ref{fr}), (\ref{me}) and  (\ref{la}) together, the time, azimuthal and radial component of the four-velocity can be expressed as
\begin{eqnarray}
  &&\dot{t}=\frac{1}{b \left[1+\frac{r^2}{2\alpha}\left(1-\sqrt{1+\frac{8\alpha M}{r^3}}\right)\right]},  \label{time} \\
  &&\dot{\phi}=\pm \frac{1}{r^2}, \label{theta} \\
  &&\dot{r}^2+\frac{1}{r^2} \left[1+\frac{r^2}{2\alpha}\left(1-\sqrt{1+\frac{8\alpha M}{r^3}}\right)\right]= \frac{1}{b^2},
 \label{radial}
\end{eqnarray}
where we have redefined the affine parameter $\lambda \rightarrow \lambda /|L|$, and $b=\frac{|L|}{E}$, which is called  the impact parameter. The  $+$ and $-$ in  Eq.(\ref{theta}) correspond to the light ray traveling in the counterclockwise and clockwise along azimuthal direction, respectively. Eq.(\ref{radial}) can also be rewritten as
\begin{equation}
 \dot{r}^2+V(r)=\frac{1}{b^2},
 \label{potential}
\end{equation}
where
\begin{equation}
V(r)=\frac{1}{r^2} \left[1+\frac{r^2}{2\alpha}\left(1-\sqrt{1+\frac{8\alpha M}{r^3}}\right)\right],
\end{equation}
is an effective potential.
The conditions for the  photon sphere orbit are $\dot{r}=0$ and $\ddot{r}=0$, which can be translated to
\begin{eqnarray}
V(r)=\frac{1}{b^2},~~~  V^{'}(r) = 0,
\label{condition1}
\end{eqnarray}
where the  prime $'$ denotes the first derivative with respect to the radial coordinate $r$. Based on this equation, we can obtain the radius $r_{ph}$ and  impact parameter $b_{ph}$ for the  photon sphere, which
 are shown together with the size of the event horizon $r_+$ in Table 1  for different $\alpha$. From this table, we can see that the three parameters, i.e., $r_{ph}$, $b_{ph}$ and $r_+$ all decrease as $\alpha$ increases.
\begin{center}
{\footnotesize{\bf Table 1.} The radius $r_{ph}$, impact parameter $b_{ph}$ of the photon sphere and the event horizon $r_+$   for different $\alpha$ with $M=1$.\\
\vspace{2mm}
\begin{tabular}{ccccccccc}
\hline       &{$\alpha=-7.7$}      & {$\alpha=-5.5$}     & {$\alpha=-3.3$} &   {$\alpha=-1.1$} &           {$\alpha=0.111$} & {$\alpha=0.333$} & {$\alpha=0.555$}  & {$\alpha=0.777$}\\    \hline
{$r_{ph}$} &  {4.70134}      &{4.36744}           &{3.95844}     &{3.40373}          &{2.94939}         &{2.83932}        &{2.71357}       &{2.56483}                  \\
{$b_{ph}$}  & {6.7815}      &{6.46004}        &{6.07084}      &{5.55557}              &{5.15252}             &{5.05903}       &{4.95501}   &{4.83671}                              \\
{$r_{+}$}  & {3.94958}     &{3.54951}         &{3.07364}       &{2.44914}              &{1.94287}             &{1.8167}       &{1.66708}   &{1.47223}                              \\
\hline
\end{tabular}} \label{tab11}
\end{center}

\begin{figure}
\centering
\subfigure[$\alpha=-5.5$]{
\includegraphics[scale=0.4]{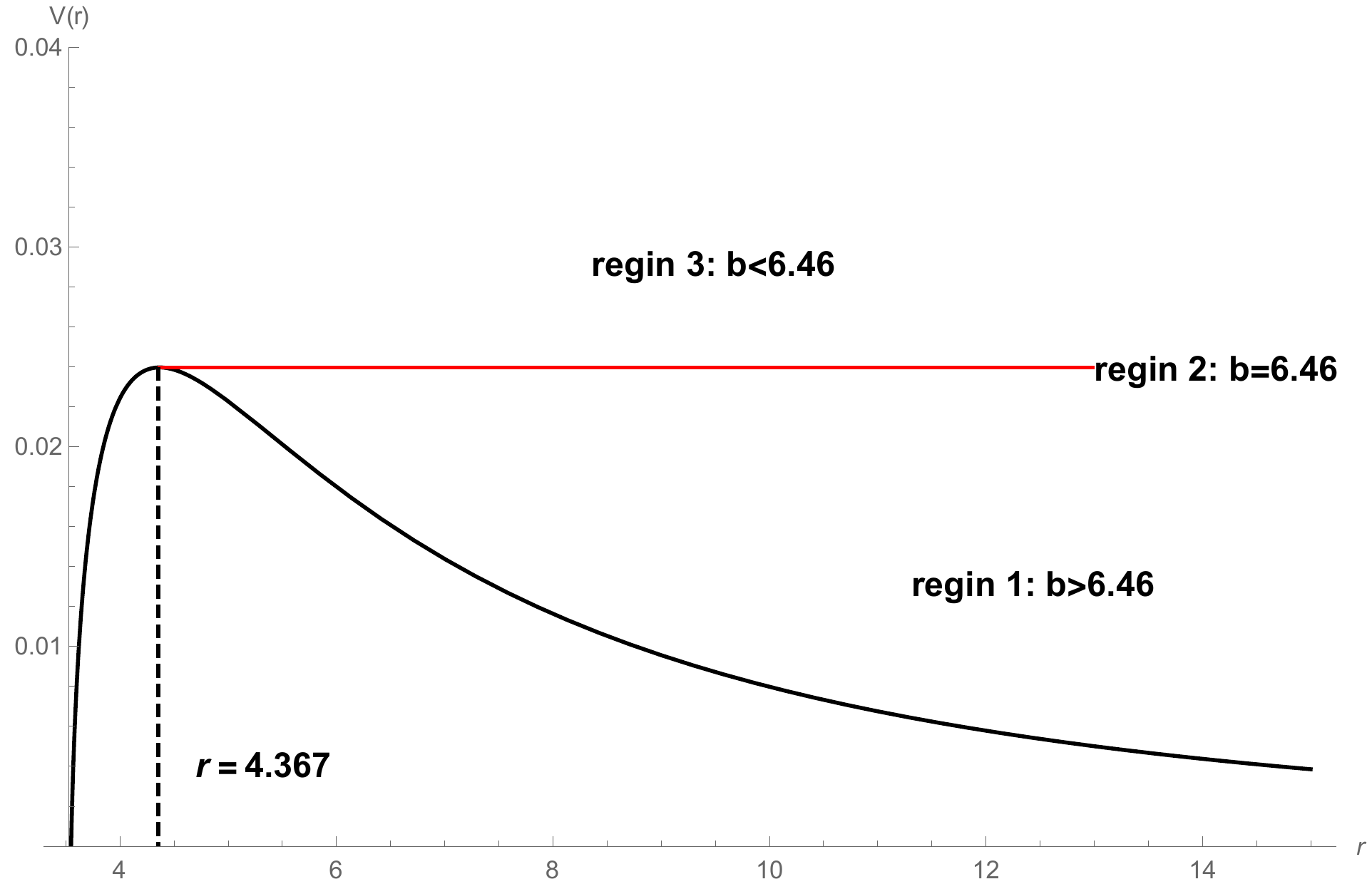}}
\subfigure[$\alpha=0.555$]{
\includegraphics[scale=0.4]{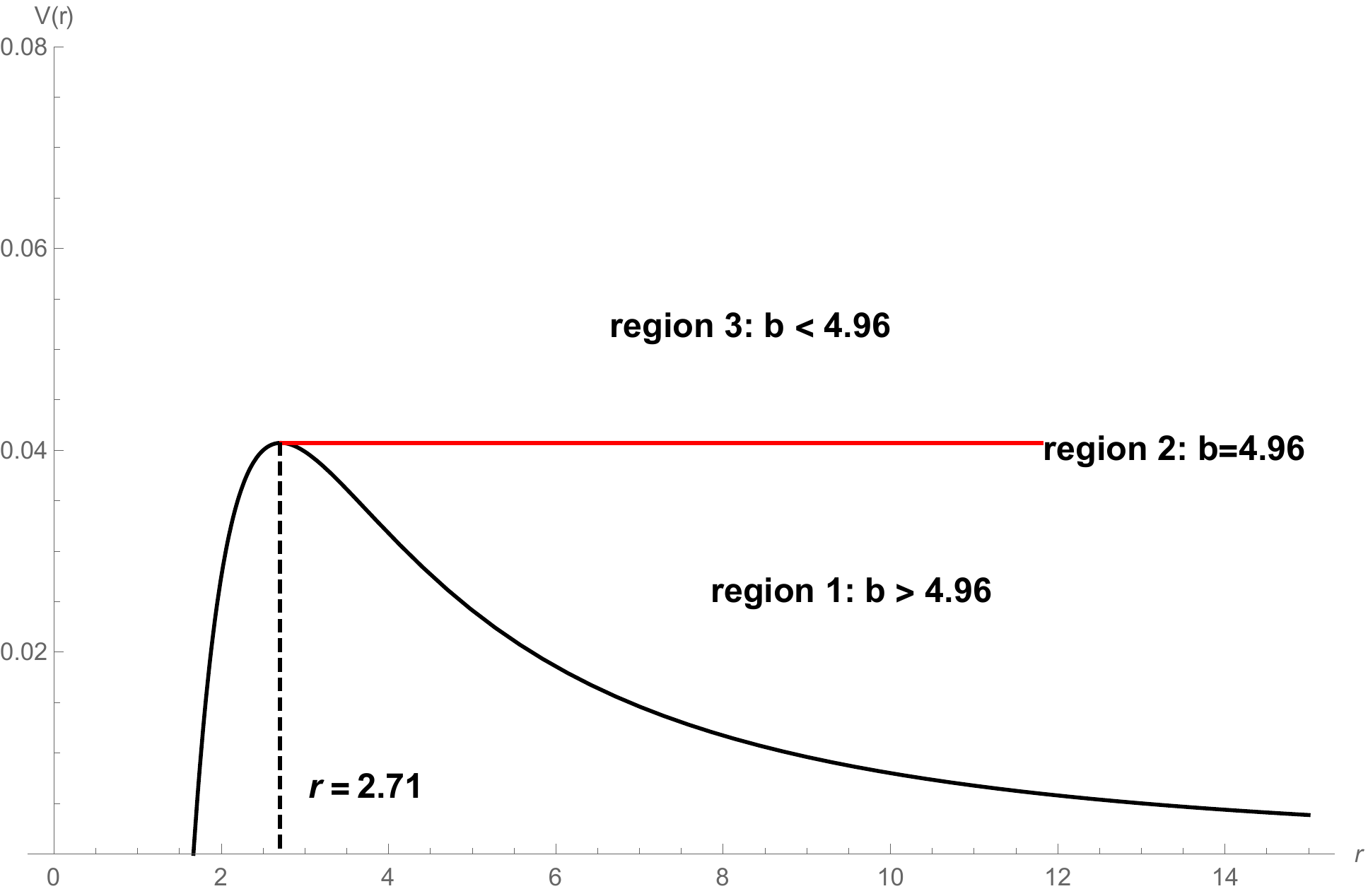}}
 \caption{\small  The profiles of the effective potential for $\alpha=-5.5$ (left panel) and $\alpha=0.555$ (right panel) with  $M=1$. For both panels, Regions 2 correspond to the red lines where $V(r)=1/b_{ph}^2$,   while  Regions  1  and  Regions 3  correspond to   $V(r)<1/b_{ph}^2$ and  $V(r)>1/b_{ph}^2$, respectively.  } \label{fig1}
\end{figure}

Here we would like to take $\alpha=-5.5, 0.555$ as two examples with the corresponding effective potential  depicted in Figure \ref{fig1}. We can see that  at the event horizon, the effective potential vanishes. It increases and reaches
a maximum at the photon sphere, and then decreases
as the light ray moves outwards.   As  a light ray moves in the radially inward direction, the effective potential will affect its  trajectory.  In Region 1, the light will encounter the potential barrier and then be reflected back in the outward direction. In Region 2, namely $b=b_{ph}$, the light will asymptotically approach the photon sphere.
 Since the angular velocity
is non-zero, it will revolve around the black hole infinitely many times. In Region 3, the light will continue moving in the inward
direction since  it does not encounter  the potential barrier. Eventually, it will enter the inside of the black hole through the event horizon.

\begin{figure}
\centering
\subfigure[$\alpha=-5.5$]{
\includegraphics[scale=0.55]{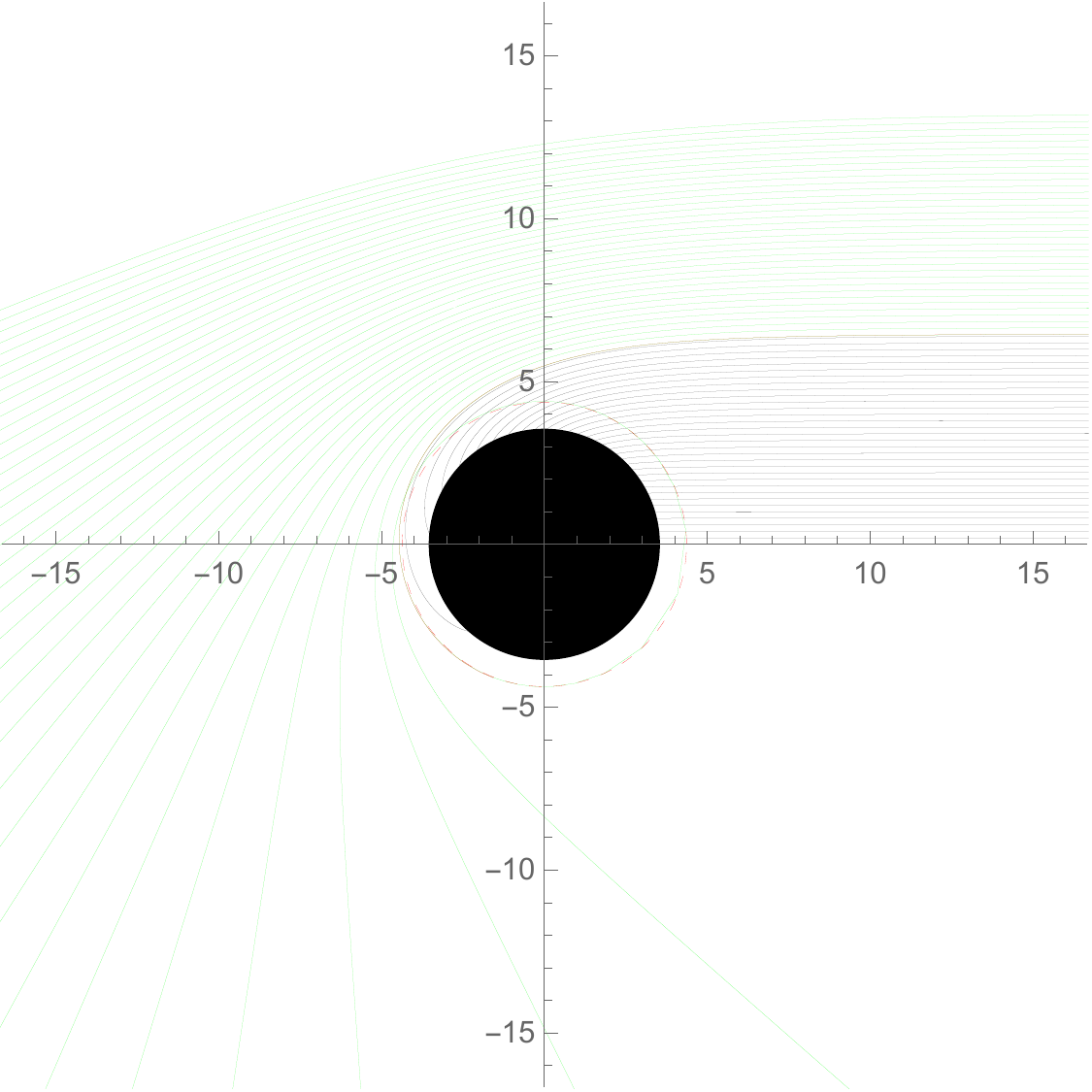}}
\subfigure[$\alpha=0.555$]{
\includegraphics[scale=0.55]{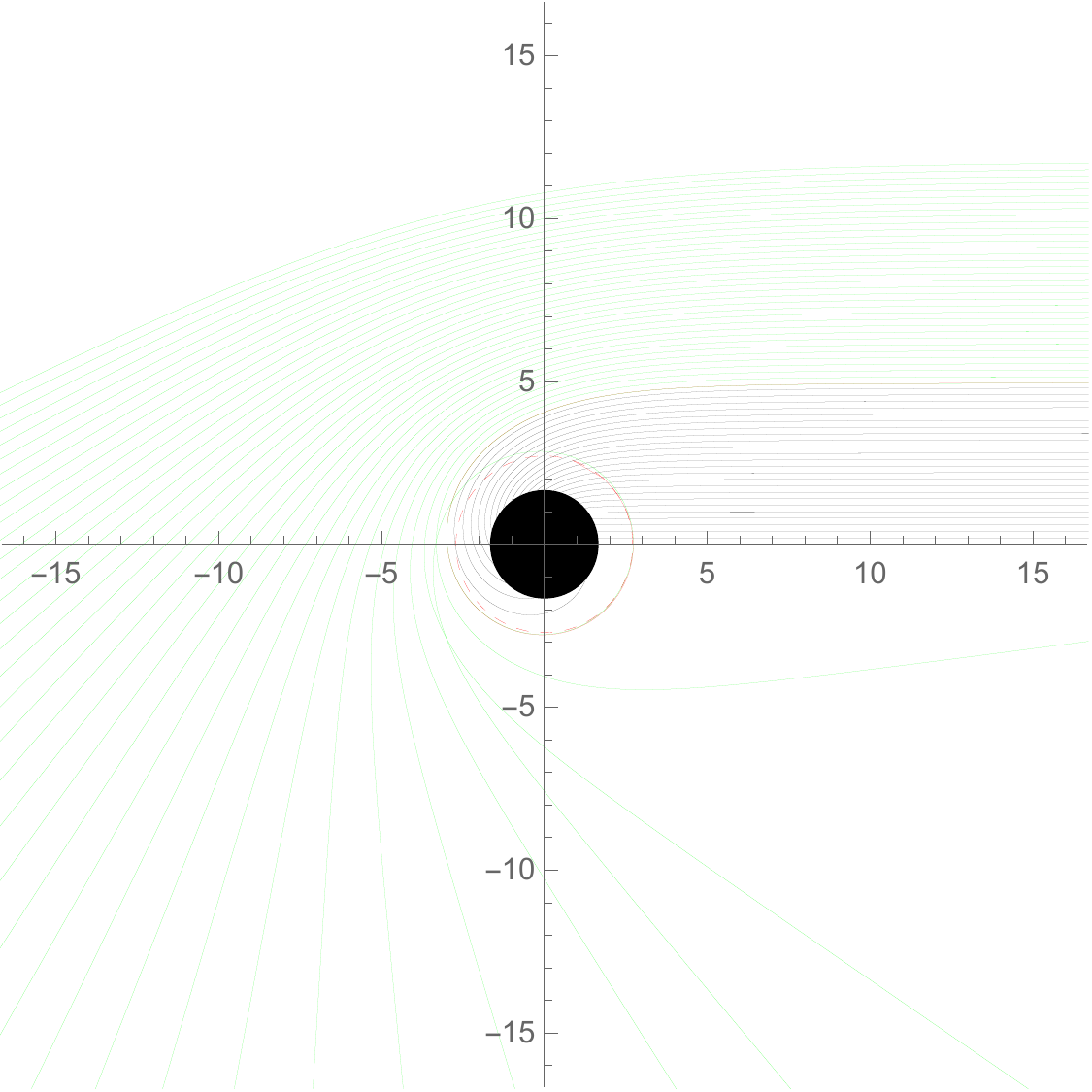}}
 \caption{\small   The   trajectory of the light ray for different $\alpha$ with   $M=1$ in the polar coordinates $(r, \phi)$.  The red line corresponds to $b=b_{ph}$, the black line corresponds to $b<b_{ph}$, and the green line  corresponds to $b>b_{ph}$. The spacing
in impact parameter is $1/5$ for all light rays. The black hole is shown as a solid disk and the photon orbit as a dashed  red  line.  }\label{fig2}
\end{figure}

Furthermore, the trajectory of the light ray can be depicted according to the equation of motion.   Combining  Eqs.(\ref{theta}) and (\ref{radial}), we have
\be
\frac{dr}{d\phi}=\pm r^2 \sqrt{\frac{1}{b^2}-\frac{1}{r^2}\left[1+\frac{r^2}{2\alpha}\left(1-\sqrt{1+\frac{8\alpha M}{r^3}}\right)\right]}. \label{drp}
\ee
By setting $u=1/r$, we can transform (\ref{drp}) into
\be
\frac{du}{d\phi}=\sqrt{\frac{1}{b^2}-u^2 \left(\frac{1-\sqrt{8 \alpha  M u^3+1}}{2 \alpha  u^2}+1\right)}\equiv G(u).\label{gu}
\ee
From Eq.(\ref{gu}) we can solve  $\phi$ with respect to $u$. Employing  the  ParametricPlot\footnote{In many references such as in Ref.\cite{Narayan:2019imo},   the ray-tracing code is employed to plot the trajectory of the light ray. }, we can plot the  trajectory of the light ray, which is shown in
Figure \ref{fig2}. The black, red and green line  correspond to $b<b_{ph}$, $b=b_{ph}$ and   $b>b_{ph}$, respectively. As one can see,
for the case of $b<b_{ph}$, the light ray drops all the way into the black hole, which corresponds to Region 3 in  Figure \ref{fig1}. For the case of $b>b_{ph}$, the light ray near  the black hole is reflected back, which corresponds to Region 1 in  Figure \ref{fig1}. And for the case of $b<b_{ph}$, the light ray revolves around the black hole, which corresponds to Region 2 in  Figure \ref{fig1}. Note that for  $b>b_{ph}$, in order to plot the geodesic, we should find a turning point, where the light ray changes   its  radial direction. The turning point is determined by the equation $ G(u)=0$, where $ G(u)$ has been defined in  Eq.(\ref{gu}).

\section{Shadows and photon spheres with  rest spherical accretion }

In this section, we will investigate the shadow  and photon sphere  of the  four-dimensional  Gauss-Bonnet black hole with static spherical accretion, which is assumed to be optically thin.
To this end, we should find   the
specific intensity observed by the observer $\rm (erg s^{-1} cm^{-2} str^{-1} Hz^{-1})$.
The observed specific intensity $I$ at the observed photon frequency $\nu_{\rm o}$
 can be found by integrating the
specific emissivity along the photon path
\cite{Jaroszynski:1997bw, Bambi:2013nla}
\be \label{intensity}
I(\nu_{\rm o}) = \int_\gamma g^3 j (\nu_{\rm e}) dl_{\rm prop} ,
\ee
where $g = \nu_{\rm o}/\nu_{\rm e}$ is the redshift factor, $\nu_{\rm  e}$ is the
photon frequency of the emitter, $dl_{\rm prop}$ is the
infinitesimal proper length, and $j (\nu_{ \rm e})$ is the
emissivity per unit volume measured in the rest frame of the emitter.
\begin{figure}[h]
\centering
\subfigure[$\alpha=-5.5$]{
\includegraphics[scale=0.5]{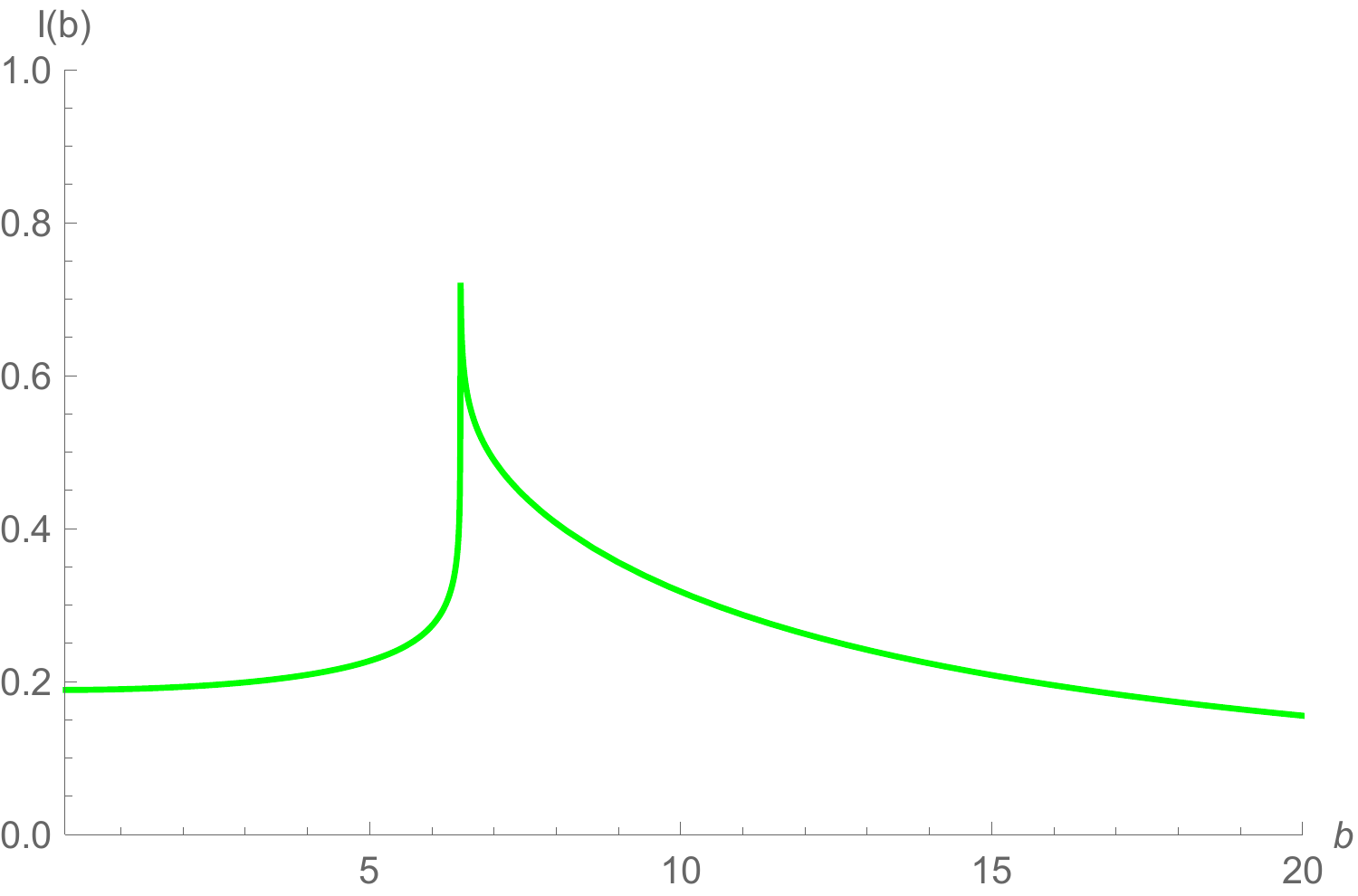}}
\subfigure[$\alpha=0.555$]{
\includegraphics[scale=0.5]{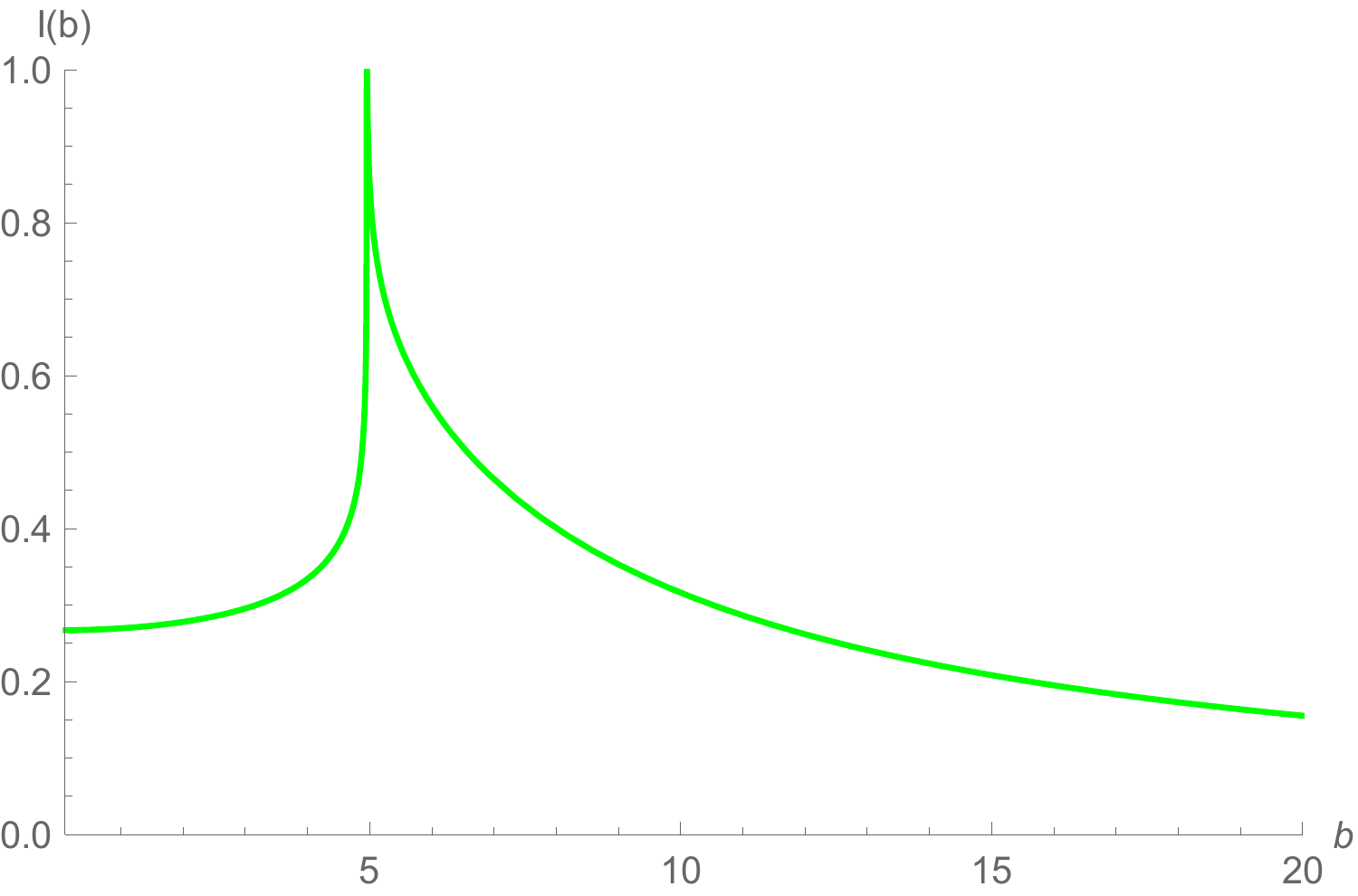}}
 \caption{\small Profiles  of  the specific intensity I(b)   seen by a distant observer for a static  spherical accretion. We set $M=1$ and take  $\alpha=-5.5$ (left panel), and $\alpha=0.555$ (right panel)  as two examples.} \label{fig3}
\end{figure}

In the four-dimensional  Gauss-Bonnet black hole,  $g =F(r)^{1/2}$.  Concerning the specific emissivity, we also
assume  that it is monochromatic with  rest-frame frequency $\nu_r$, that is
\begin{equation}
j (\nu_{\rm e})\propto \frac{\delta(\nu_e-\nu_r)}{r^2}.  \label{profile}
\end{equation}
  According to  Eq.(\ref{metric}), the  proper length measured in the rest frame of the emitter is
\bea
dl_{\rm prop}&=&\sqrt{F(r)^{-1}dr^2+r^2 d \phi^2}\nonumber \\
  &=&\sqrt{F(r)^{-1}+r^2 (\frac {d \phi}{dr})^2} dr,
\ena
in which $ d \phi/dr$ is given by Eq.(\ref{drp}). In this case, the specific intensity observed by the infinite observer is
\be \label{finalintensity}
I(\nu_{\rm o}) = \int_\gamma \frac{F(r)^{3/2}} {  r^2} \sqrt{F(r)^{-1}+r^2 (\frac {d \phi}{dr})^2} dr.
\ee
The intensity is circularly symmetric, with the impact parameter $b$ of the radius, which satisfies $b^2=x^2+y^2$.

Next we will employ Eq.(\ref{finalintensity}) to investigate the shadow of the  four-dimensional  Gauss-Bonnet black hole with the static spherical accretion. Note that the intensity depends on the
trajectory of the light ray, which is determined by the impact parameter $b$. So we will investigate how the  intensity  varies with respect to the impact parameter. For different $\alpha$, the numerical results of $I(b)$ are shown in Figure \ref{fig3}. From this figure, we see that the  intensity increases rapidly  and reaches a
peak at  $b_{ph}$, and then drops to lower values  with increasing $b$. This result is consistent with Figure \ref{fig1} and  Figure \ref{fig2}. Since for   $b<b_{ph}$, the intensity originating from the accretion is absorbed   mostly by the black hole. And for $b=b_{ph}$, the light ray revolves around the black hole many times, so the observed intensity is maximal. While for $b>b_{ph}$, only the refracted light contributes to the   intensity of the observer. As $b$ becomes larger, the refracted light becomes less. The observed  intensity thus vanishes for large enough  $b$. In principle, the peak intensity at $b = b_{ph}$  should be infinite because the light ray    revolves around the black hole infinite times  and
collect an arbitrarily large intensity.  However, because of the numerical limitations and the logarithmic form of the intensity, the real computed intensity never goes
to infinity, which has also been well addressed in \cite{Gralla:2019xty, Narayan:2019imo}.
From Figure \ref{fig3}, we can also observe how the  Gauss-Bonnet  coupling
constant affects the observed intensity. For all the $b$,  the larger the coupling
constant is, the stronger the intensity will be.

\begin{figure}[h]
\centering
\subfigure[$\alpha=-5.5$]{
\includegraphics[scale=0.5]{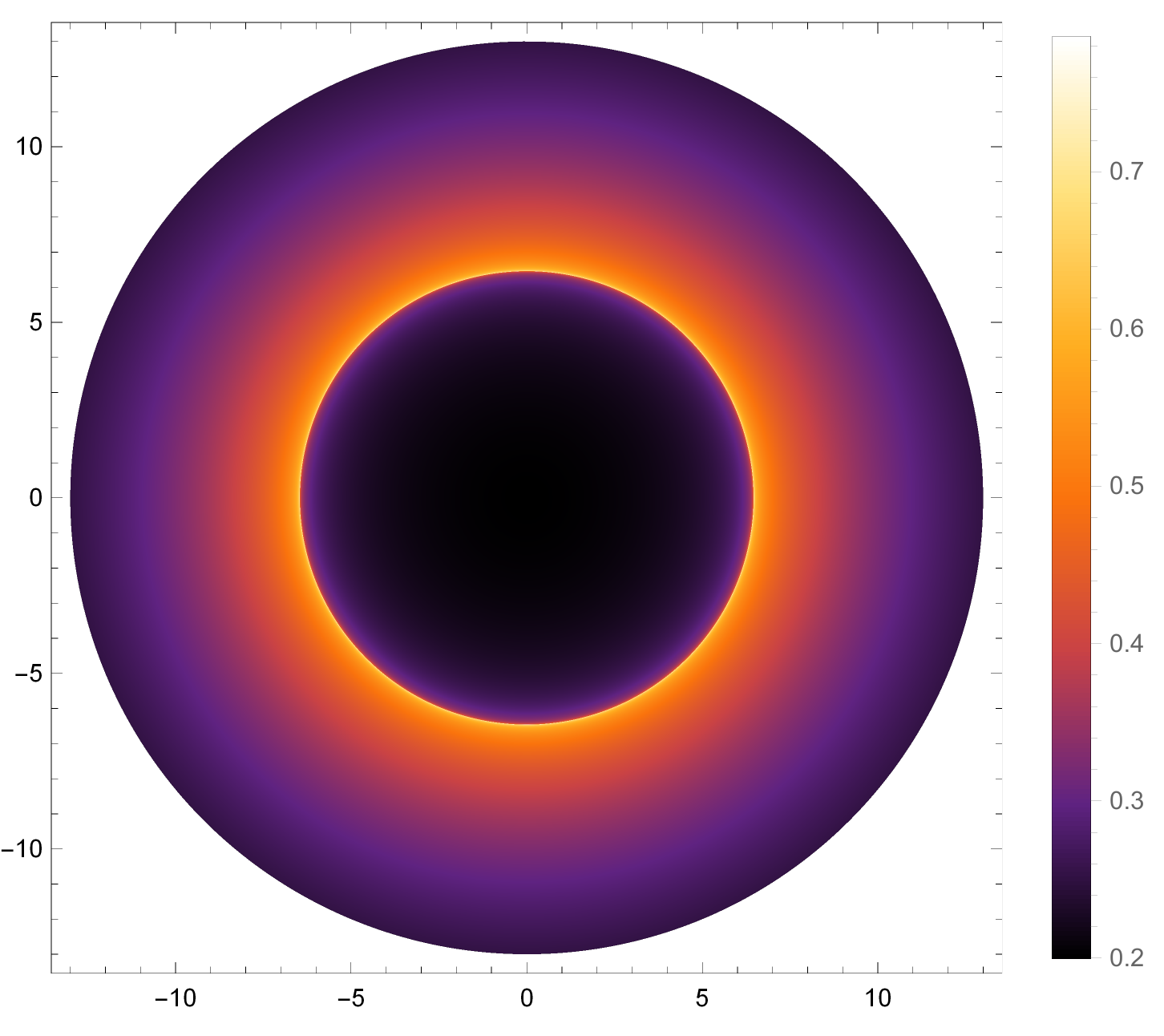}}
\subfigure[$\alpha=0.555$]{
\includegraphics[scale=0.5]{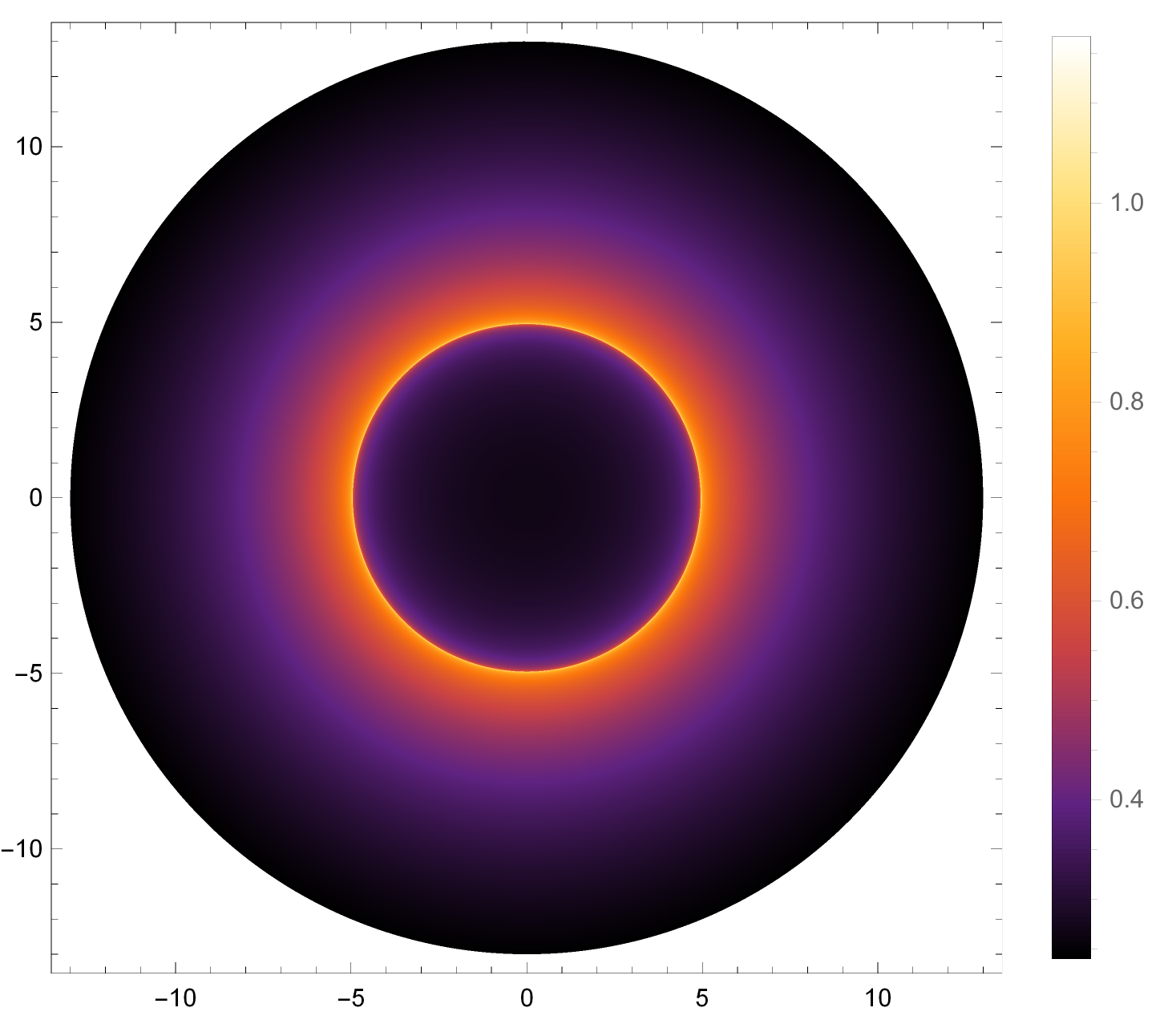}}
 \caption{\small  The black hole shadows  cast by the static accretion   for different $\alpha$ with   $M=1$ in the $(x ,y)$ plane. The bright ring is the photon sphere.  } \label{fig4}
\end{figure}

The shadow  cast by the  four-dimensional  Gauss-Bonnet black hole in the  $(x ,y)$ plane is shown in Figure \ref{fig4}. We can see that outside the black hole shadow, there is a bright ring, which is the photon sphere.
The radii of the photon spheres for different $\alpha$ have been listed in Table 1. Obviously, the results in Figure \ref{fig4} are consistent with those in Table 1. That is, the larger the Gauss-Bonnet
constant is, the smaller the  radius of the photon sphere will be.

Moreover, we can see that inside
the shadow, the intensity does not go to zero
but having a small finite value. The reason is that part of the  radiation  has escaped
to infinity. For $r>r_{ph}$, the solid angle of the
escaping rays is  $2 \pi (1+\cos\theta)$, while for $r<r_{ph}$, the solid angle of the
escaping rays is  $2 \pi (1-\cos\theta)$, where $\theta$  is given by
\be
\sin \theta =\frac{r_{ph}^{3/2}}{r} \left[ 1+\frac{r^2}{2\alpha}\left(1-\sqrt{1+\frac{8\alpha M}{r^3}}\right) \right]^{1/2}.
\ee
By only counting the escaping light rays, we have the net luminosity observed at infinity as

\be
L_{\infty}=\int_{r_+}^{r_{ph}} 4 \pi r^2 j (\nu_{\rm e}) 2 \pi (1-\cos\theta) dr+\int_{r_{ph}}^{\infty} 4 \pi r^2 j (\nu_{\rm e}) 2 \pi (1+\cos\theta) dr.
\ee

\begin{center}
{\footnotesize{\bf Table 2.} The net luminosity of the escaping rays for different $\alpha$ with $M=1$.\\
\vspace{2mm}
\begin{tabular}{ccccccccc}
\hline       &{$\alpha=-7.7$}      & {$\alpha=-5.5$}     & {$\alpha=-3.3$} &   {$\alpha=-1.1$} &           {$\alpha=0.111$} & {$\alpha=0.333$} & {$\alpha=0.555$}  & {$\alpha=0.777$}\\    \hline
{$L_{\infty}$} &  {0.169698}      &{0.18729}           &{0.215428}     &{0.265181}          &{0.323863}         &{0.34006}        &{0.359979}       &{0.386447}                  \\
\hline
\end{tabular}} \label{tab2}
\end{center}
For different $\alpha$, the numerical results are listed in Table 2. We can see that the net luminosity increases with increasing $\alpha$.  For the Schwarzschild black hole, the net luminosity is found to be $L_{\infty}=0.32$  \cite{Narayan:2019imo}.
Obviously, for the positive  $\alpha$, the net luminosity in the  four-dimensional  Gauss-Bonnet black hole  is larger than that in Schwarzschild black hole, while for the negative  $\alpha$, the net luminosity in this spacetime is smaller than  that in Schwarzschild black hole.

\section{Shadows and photon spheres with infalling  spherical accretion }

In this section, we  allow the optically thin accretion  to
move  towards the black hole. This model is thought to be more realistic than the static accretion model since most of  the accretions are mobile in the universe.
For simplicity, we assume  that the accretion free falls on
to the black hole from infinity.  We still employ  Eq.(\ref{finalintensity}) to investigate the shadow of the  four-dimensional  Gauss-Bonnet black hole.

Different from the static accretion, the
redshift factor for the infalling accretion  should be  evaluated by
\begin{equation}
g = \frac{k_\beta u^\beta_{\rm o}}{k_\gamma u^\gamma_{\rm e}} , \label{redf}
\end{equation}
in which  $k^\mu=\dot{x_\mu}$ is the four-velocity of the photon, $u^\mu_{\rm o} = (1,0,0,0)$ is
the 4-velocity of the distant observer, and $u^\mu_{\rm e}$ is the 4-velocity of
the accretion under consideration, given by
\bea
u^t_{\rm e} = \frac{1}{F(r)},~~~
u^r_{\rm e} = - \sqrt{ 1 - F(r) } ,~~~
u^\theta_{\rm e} = u^\phi_{\rm e} = 0.
\ena

The four-velocity  of the photon has been  obtained
previously in  Eq.(\ref{time})- Eq.(\ref{radial}). We know that  $k_t=1/b$ is a constant, and $k_r$ can be inferred
from $k_\gamma k^\gamma = 0$. Therefore,
\begin{equation}
\frac{k_r}{k_t} = \pm \frac{1}{F(r)}\sqrt{   1 - \frac{b^2 F(r)}{r^2} } ,
\end{equation}
where the sign $+$($-$) correspond to the case that the photon gets close to (away from)
the black hole. With this equation, the redshift factor in Eq.(\ref{redf}) can be simplified as
\begin{equation}
g =  \frac{1}{u^{t}_e+k_r/k_e u^r_e}.\label{sredf}
\end{equation}
In addition, the proper  distance can be defined as
\begin{equation}
dl_{\rm prop} = k_\gamma u^\gamma_{\rm e}
d\lambda=\frac{k_t}{g | k_r |} dr,
\end{equation}
where $ \lambda$ is the
affine parameter along the photon path $\gamma$.  We also
assume that the specific emissivity is monochromatic, therefore, Eq.(\ref{profile})  can be used.
The intensity in Eq.(\ref{intensity})  thus can be expressed as
\bea
I(\nu_o) \propto  \int_\gamma \frac{g^3 k_t dr}{r^2 | k_r |}  .   \label{finaltensity}
\ena

\begin{figure}[h]
\centering
\subfigure[$\alpha=-5.5$]{
\includegraphics[scale=0.5]{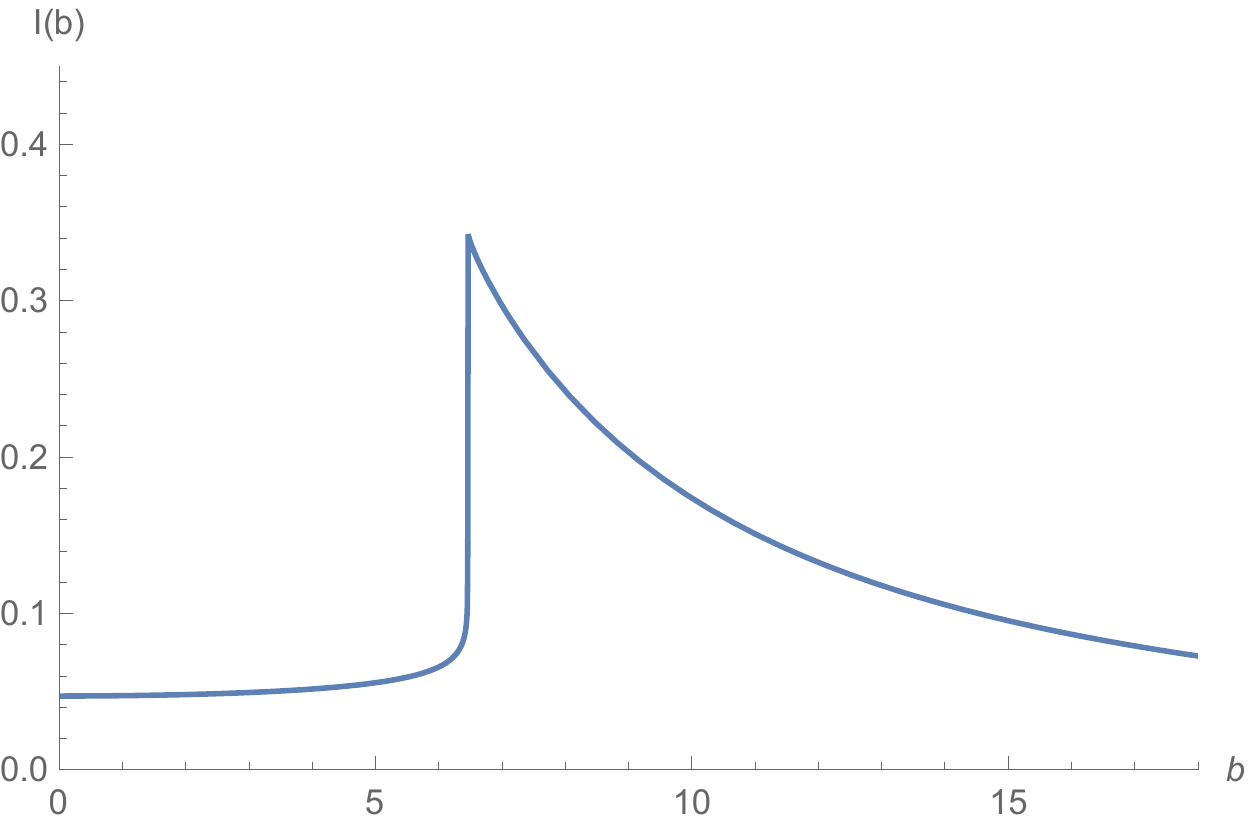}}
\subfigure[$\alpha=0.555$]{
\includegraphics[scale=0.5]{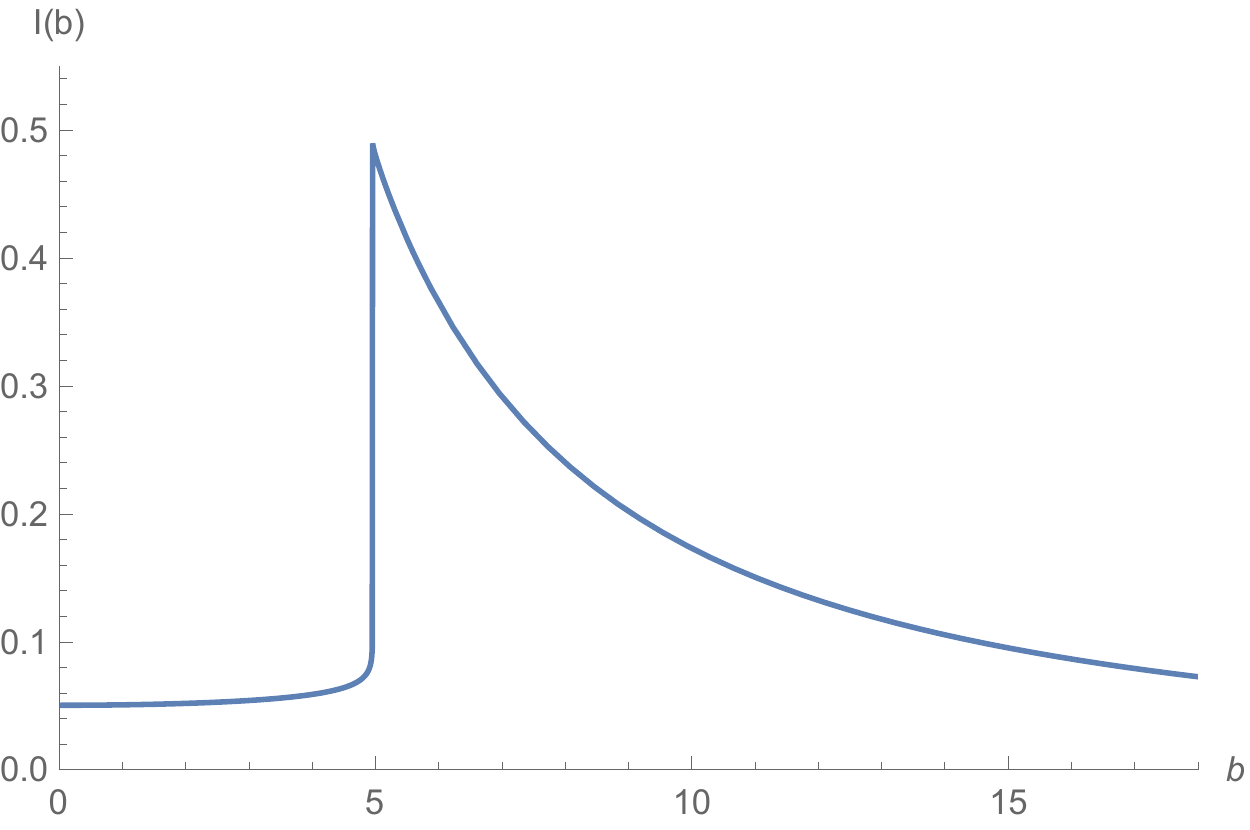}}
 \caption{\small  The profiles  of  the specific intensity I(b)  seen by a distant observer for an infalling  accretion. For both cases, we set  $M=1$. } \label{fig5}
\end{figure}

Now we will use  Eq.(\ref{finaltensity}) to investigate the shadow of the black hole numerically. For different $\alpha$, the intensities with respect to $b$ observed by the distant observer are  shown in Figure \ref{fig5}.
Similar to the static accretion, we find that as $b$ increases, the intensity increases first,
then reaches a maximum intensity at $b = b_{ph}$, and
then drops away. We can also observe the effect of $\alpha$ on the intensity from Figure \ref{fig5}. That is, the larger the value of  $ b$ is, the larger the observed intensity will be.

\begin{figure}[!h]
\centering
\subfigure[$\alpha=-5.5$]{
\includegraphics[scale=0.5]{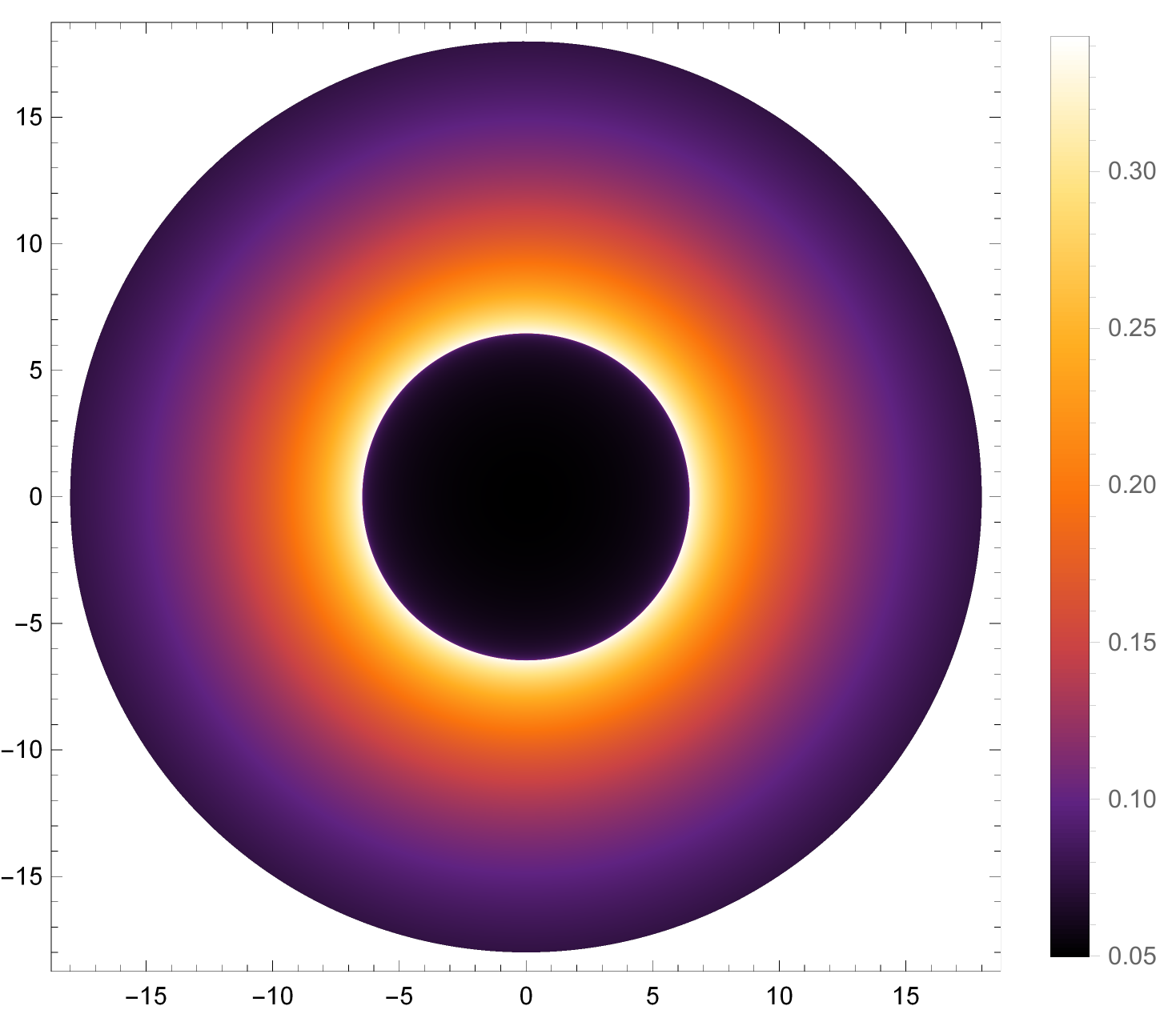}}
\subfigure[$\alpha=0.555$]{
\includegraphics[scale=0.5]{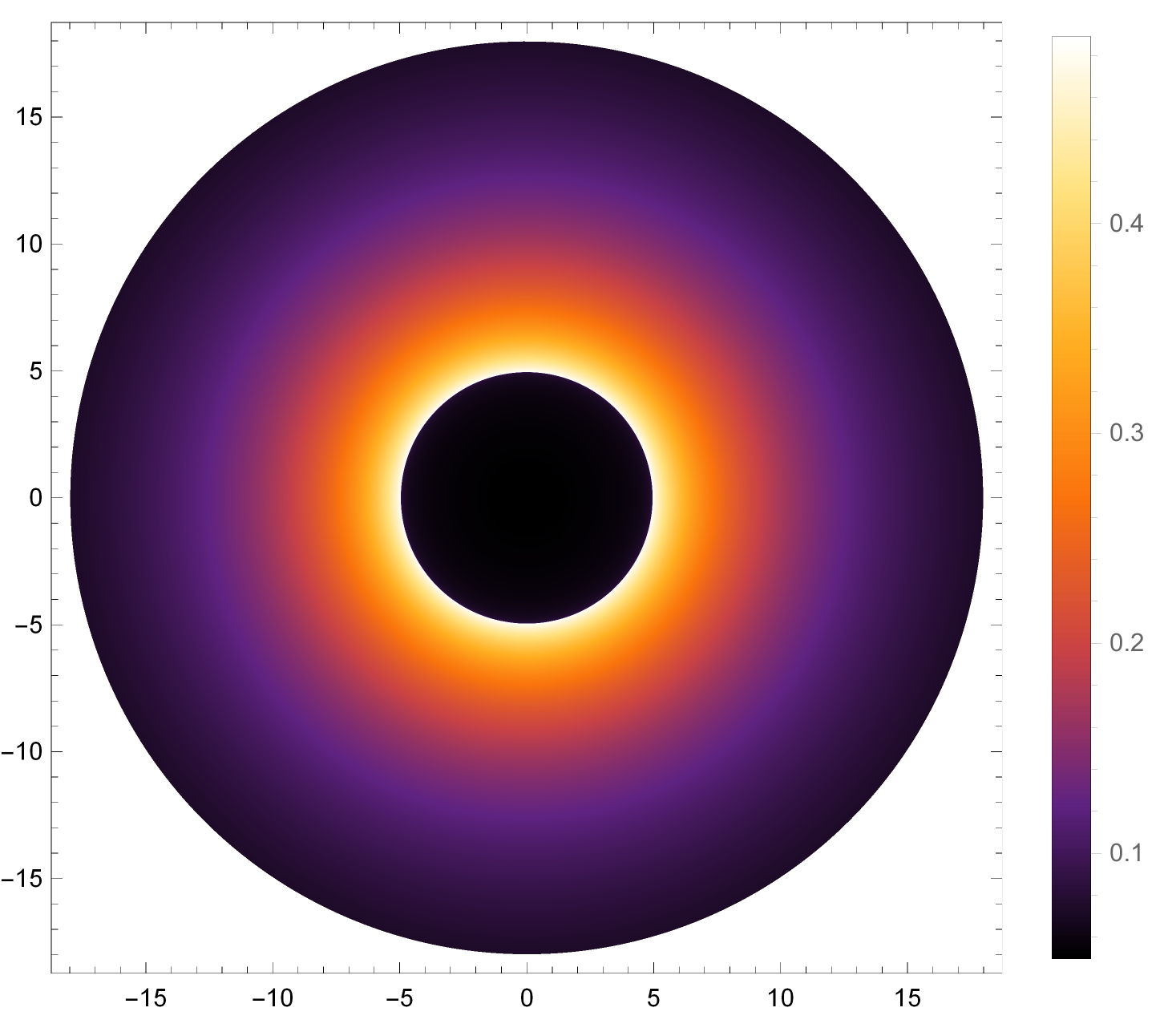}}
 \caption{\small The shadow of the black hole  cast by the infalling  accretion   for different $\alpha$ with   $M=1$ in the $(x ,y)$ plane.  The brightest  ring outside the black hole is the photon sphere. } \label{fig6}
\end{figure}

The 2-dimensional  image of the shadow and photon sphere seen by a distant observer are shown in
Figure \ref{fig6}. We can see that the radius of the shadow and the location of the photon sphere are the same as those with the static accretion. A
major new feature is that in the central region, the shadow with infalling accretion is darker than that with the static accretion, which is well accounted for by the Doppler effect.  Nearer the event horizon of the black hole, this effect is more obvious.

It has been argued that in the universe, the accretion flows
do have inward radial velocity, and the velocity tends
to be large precisely at the radii of interest for the shadow
formation. Therefore the model with radially infalling gas is
most appropriate for comparison with the image of M87*.

\begin{figure}[h]
\centering
\subfigure[$ j(\nu_{\rm e})=1/r^4$]{
\includegraphics[scale=0.5]{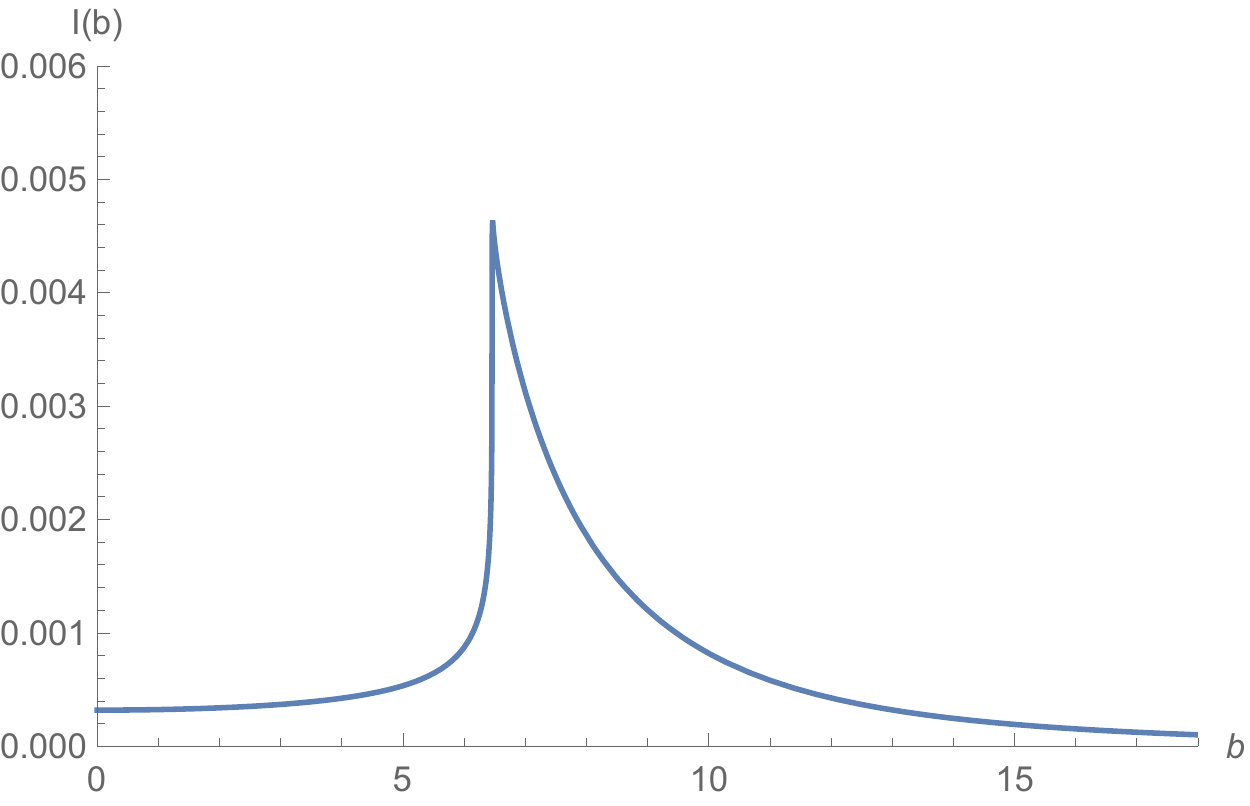}}
\subfigure[$ j(\nu_{\rm e})=1/r^5$]{
\includegraphics[scale=0.5]{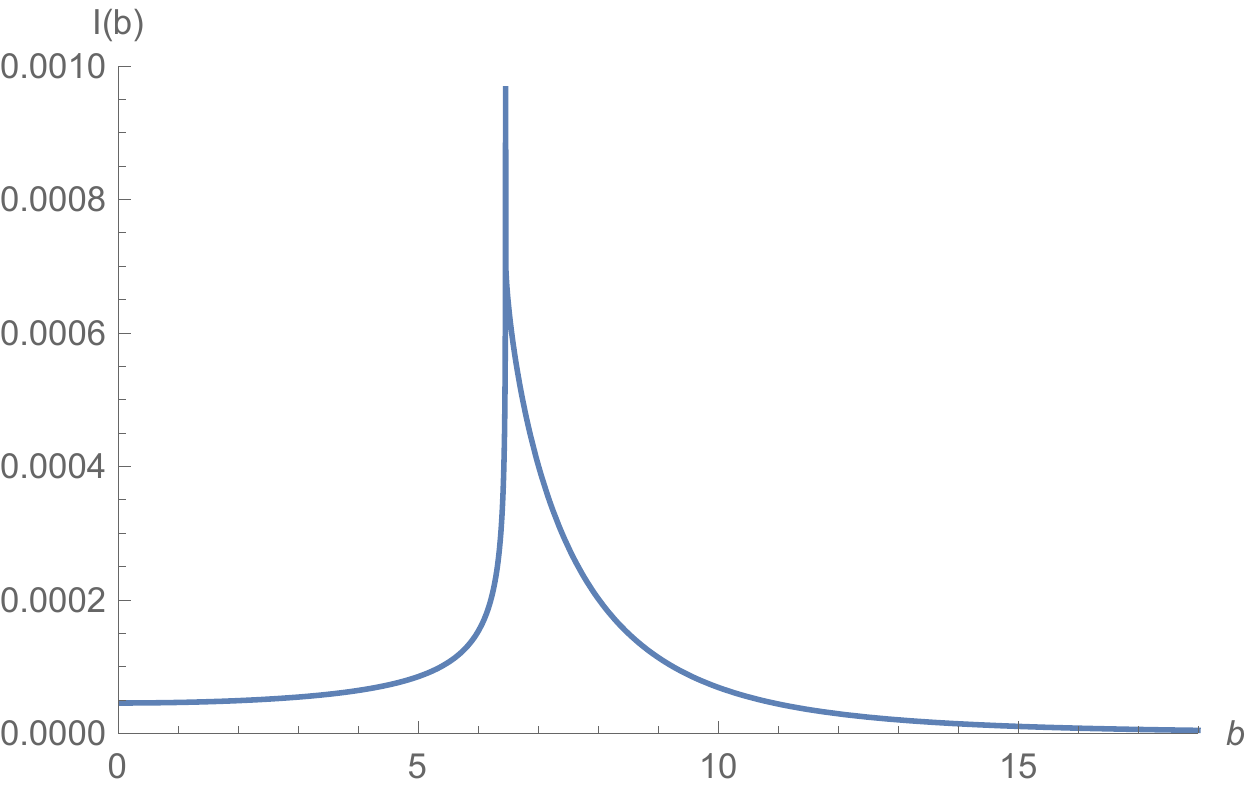}}
 \caption{\small  Profile  of  the specific intensity I(b)   seen by a distant observer with different profiles of specific emissivity. For both cases, we set  $M=1$, $\alpha=-5.5$. } \label{fig7}
\end{figure}

\begin{figure}[h]
\centering
\subfigure[$j(\nu_{\rm e})=1/r^4$]{
\includegraphics[scale=0.5]{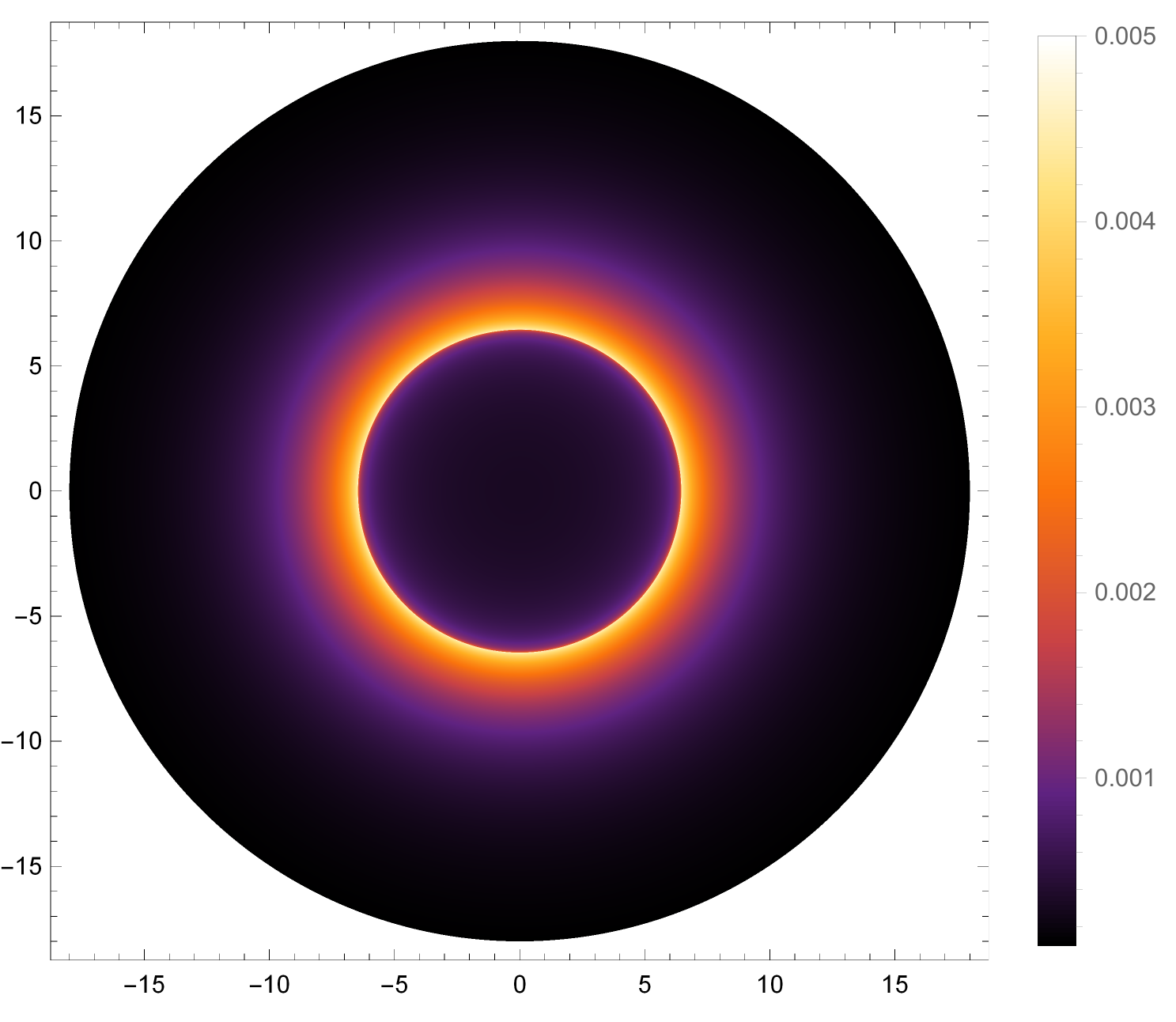}}
\subfigure[$ j(\nu_{\rm e})=1/r^5$]{
\includegraphics[scale=0.5]{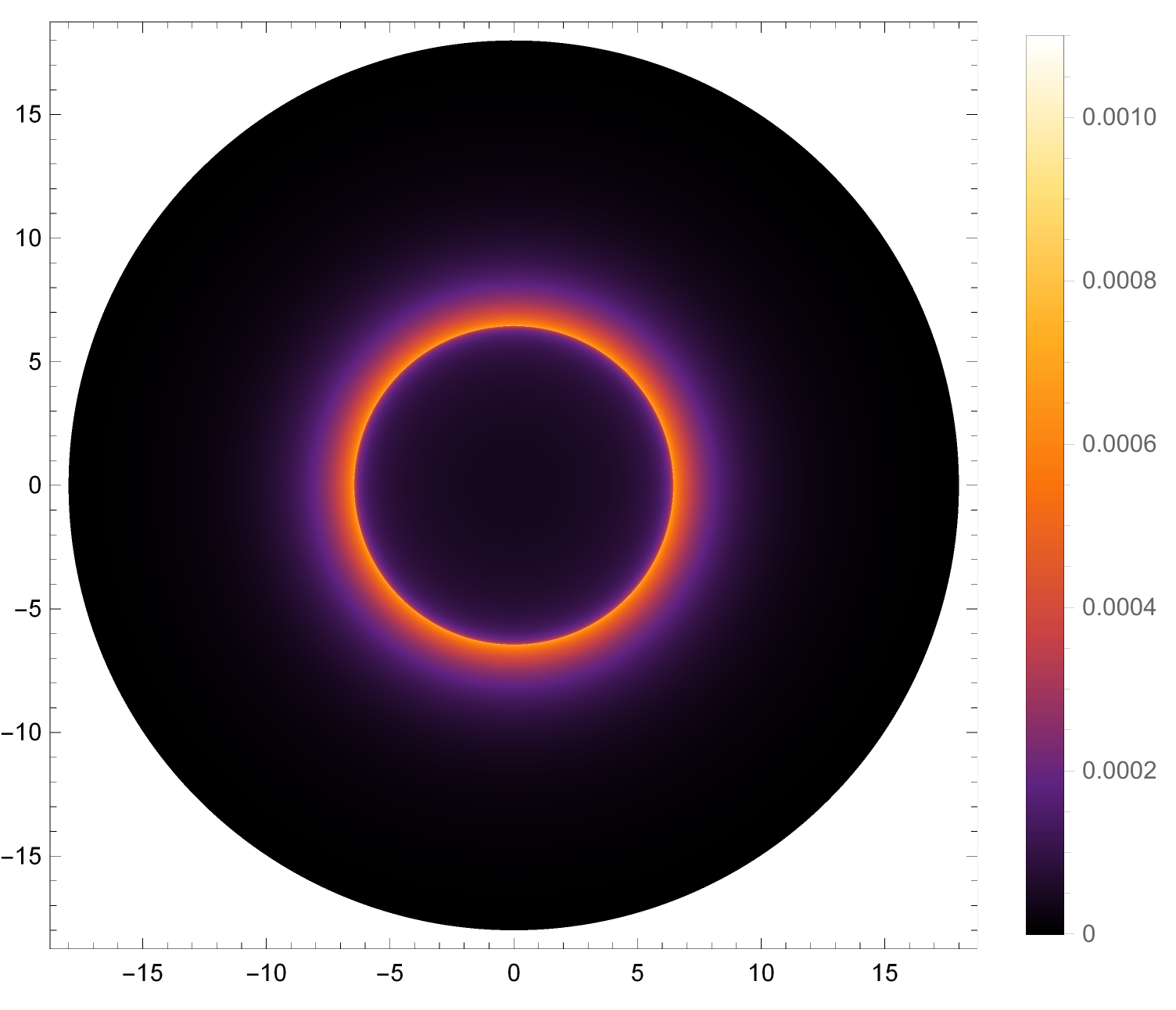}}
 \caption{\small   The shadow of the black hole cast by the infalling  accretion with different profiles of specific emissivity  in the $(x ,y)$ plane. We set $\alpha=-5.5$, $M=1$. } \label{fig8}
\end{figure}

In order to explore how the  profile of the specific emissivity  affects the shadow of the black hole, we will choose different profiles of $ j(\nu_{\rm e})$. The corresponding  intensities are shown in Figure \ref{fig7}.
From  this figure, we see clearly that the intensity in these cases has the behavior similar to the case  $ j(\nu_{\rm e})=1/r^2$. That is, the peak is always located at $b=b_{ph}$. The difference is that the
intensity  decays  faster for the higher power of $1/r$, which makes the peak more prominent. The corresponding 2-dimensional image of shadow and photon sphere are shown in Figure \ref{fig8}.

Our results in Figure \ref{fig7} and Figure \ref{fig8} show that although  the profile of the  spherical  accretion affects the intensity of the shadow, it does not affect the characteristic geometry such as the radius of the shadow, which is determined  only by the geometry of the spacetime.

\section{Discussions and conclusions }

In this paper, we have investigated the shadows and photon spheres cast by the four-dimensional  Gauss-Bonnet black hole with spherical accretions. We first obtain the radius of the photon sphere and critical impact parameter for different Gauss-Bonnet constants, and find that the  larger the Gauss-Bonnet constant is, the smaller the  radius of the photon sphere and critical impact parameter will be, which is consistent with the previous results \cite{Guo:2020zmf, Wei:2020ght}. It should be noted that, a simple approximation of the radius of the shadow was derived in Section V of \cite{Konoplya:2020bxa}, in which the authors mainly studied the quasi-normal modes and stability of the  four dimensional spherical Gauss-Bonnet black hole. In the concise Section V of \cite{Konoplya:2020bxa} the authors analytically obtained a linear relation between the Gauss-Bonnet constant $\alpha$ with respect to the radius of the shadow, in the units of event horizon. This linear relation indeed was only satisfied in the small $\alpha$ regime. In fact, from the numerics in Table 1 in our paper we can check that in the units of event horizon, the radius of the shadow also increase as $\alpha$ grows. However, for larger $\alpha$'s this linear relation will be destroyed. Therefore, our numerical evaluations actually go beyond the simple derivations in \cite{Konoplya:2020bxa}.

More importantly, we obtain the specific intensity $I(\nu_o)$ observed by a distant observer, in which the accretion was supposed to be either static or infalling. For both cases, we find that the specific intensity
increases with the increasing Gauss-Bonnet constant. We plot the image of the shadows in the  $(x, y)$ plane, and find that there is a bright sphere ring outside the dark region. The interior region of the shadow with the infalling accretion turns out to be darker than that with the static accretion, due to the Doppler effect. We also investigate the effect of the profile of the accretion on the shadow. As a result, it is found that  although  the profile will  affect the intensity of shadow, it does not affect the characteristic of the geometry such as the radius of the photon sphere. In  Ref.\cite{Narayan:2019imo}, the emission originating from the accretion was  cut-off at different locations, the size of the shadow was found to be independent of the locations.  Obviously, our result is consistent with the observation in  Ref.\cite{Narayan:2019imo}.

The EHT Collaboration has molded M87* with the Kerr black hole, and claimed that the observation supports the General Relativity. In this paper, we did not consider   the Kerr-like black hole in the four-dimensional Gauss-Bonnet gravity since  the spherically symmetric black hole,  in some case,  may produce qualitatively similar results \cite{Jaroszynski:1997bw}.  For example,
the simplified spherical model captures the key features that also appear in state of the art general-relativistic
magnetohydrodynamics models \cite{Akiyama:2019fyp}, whether they are spinning or not.

\begin{figure}[h]
\centering
\subfigure {
\includegraphics[scale=0.55]{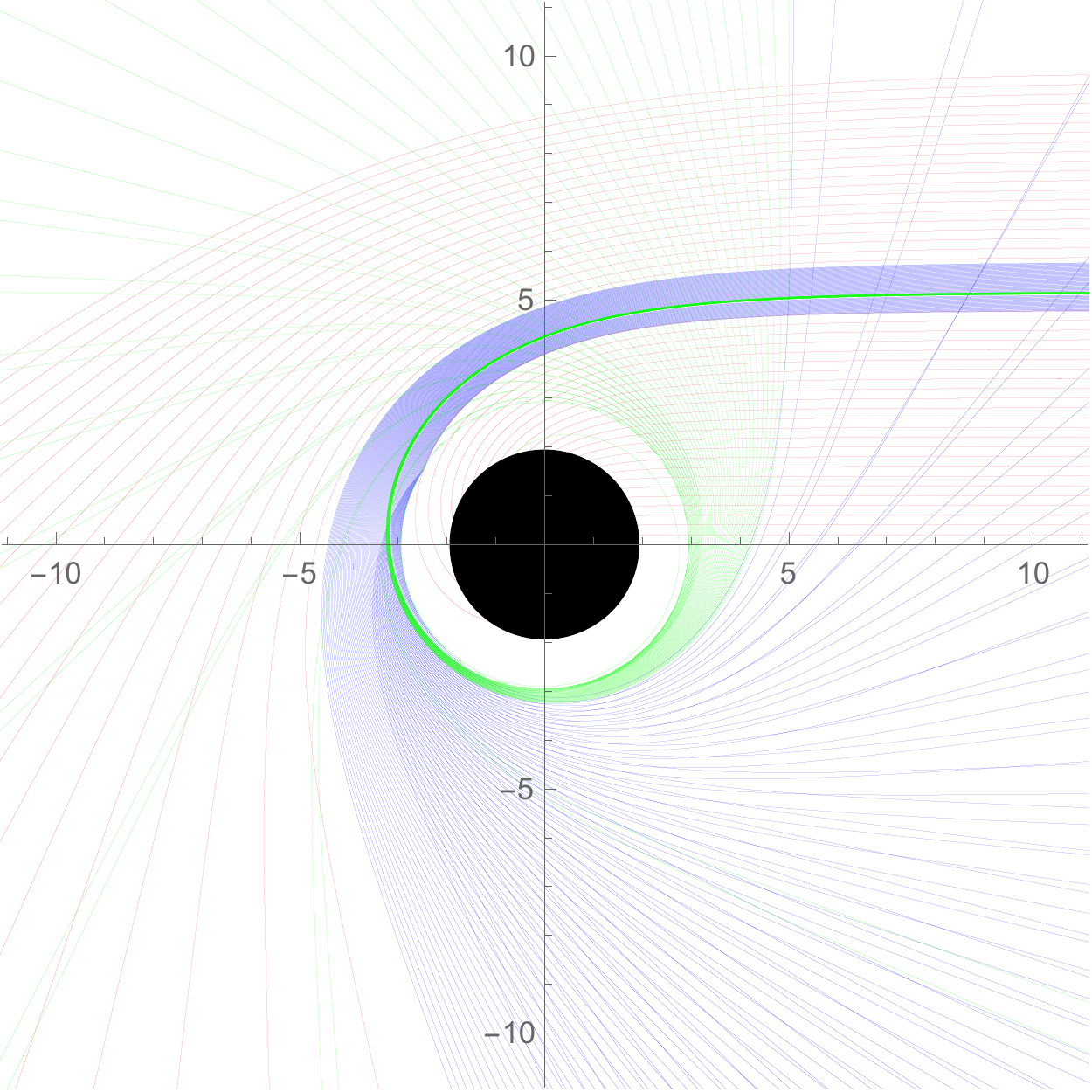}}
 \caption{\small  The behavior of light rays  as a function of impact parameter $b$. We treat $(r, \phi)$ as the Euclidean polar coordinates. The red lines, blue lines and green lines correspond to the direct, lensed, and photon ring trajectories, respectively. The spacing in impact parameter is $1/5, 1/100,  1/1000$ in the direct, lensed, and photon ring bands. The black hole is
shown as a solid disk and the photon orbit as a dashed line. We set $\alpha=-5.5$, $M=1$. } \label{fig9}
\end{figure}

In addition, the real accretion flows are generically not spherically symmetric.
The hot accretion
flow in   M87* and most
other galactic nuclei consist of a geometrically thick and quasi-spherical disk. It will be more interesting to investigate the shadow with a thick disk accretion. Recently,  Ref.\cite{Gralla:2019xty} has investigated the shadow with a thin and thick accretion. They reanalyzed the orbit of photon and redefined the photon ring and lensing ring, in which the lensing ring is the  light ray that intersects the
plane of the disk twice and  the photon ring is that  intersects the plane
three or more times. They defined a total number
of orbits as $n\equiv \phi/2 \pi$.  In this case,    $n>3/4$ corresponds to the light ray
crossing the equatorial plane at least twice, $n>5/4$  corresponds to the light ray  crossing the
equatorial plane at least three times, and $n<3/4$ corresponds to  the light ray   crossing the
equatorial plane only once. For the case of $\alpha=-5.5$, the trajectory of the light ray is shown in Figure \ref{fig9}. Compared it with  Figure \ref{fig2}, we see that the photon ring is around the photon sphere, and the lensing ring is around the photon ring.  It will be interesting to investigate the shadow, photon ring, and lensing ring with a thin or  thick  disk in the four-dimensional  Gauss-Bonnet black hole. We leave it as future work.

\section*{Acknowledgements}{We are grateful to Xiaoyi Liu  for her invaluable discussions throughout this project. This work is supported  by the National
Natural Science Foundation of China (Grant Nos. 11875095,  11675015, 11675140, 11705005). In addition, H.Z.
is supported in part by FWO-Vlaanderen through the
project G006918N, and by the Vrije Universiteit Brussel
through the Strategic Research Program ``High-Energy
Physics''. He is also an individual FWO fellow supported
by 12G3518N.}

\end{document}